\documentclass[12pt]{article}

\usepackage{graphicx}
\usepackage{amsmath}

\newcommand{\beq}{\begin{equation}}
\newcommand{\eeq}{\end{equation}}
\newcommand{\beqa}{\begin{eqnarray}}
\newcommand{\eeqa}{\end{eqnarray}}
\newcommand{\beqar}{\begin{eqnarray*}}
\newcommand{\eeqar}{\end{eqnarray*}}

\begin{document}
\thispagestyle{empty}

\vspace{32pt}
\begin{center}
{\textbf{\Large The Geometric Origin of the CP Phase}}
\vspace{40pt}

H.~Fanchiotti$^1$, C.A.~Garc\'\i a Canal$^1$, V.~Vento$^2$\\
 \vspace{12pt}
\textit{$^1$IFLP/CONICET and Departamento de F{\'\i}sica}\\ \textit{Universidad Nacional de La Plata, C.C.67, 1900, La Plata, Argentina}\\
\vspace{10pt} \textit{$^2$Departamento de F{\'\i}sica Te\'orica-IFIC. Universidad de Valencia-CSIC.}\\ \textit{E-46100, Burjassot (Valencia), Spain}

\end{center}

\vspace{40pt}

\date{\today}

\begin{abstract}

The complex phase present in CP-violating systems such as neutral kaons is
shown to be of geometrical origin. It is also concluded that the complex phase
of the Cabibbo--Kobayashi--Maskawa (CKM) matrix is a Berry-like phase.
\end{abstract}

\newpage

Following the original analogy presented in Ref.\ \cite{Rosner} between the
physics of the weak decay of neutral K-mesons (kaons) and a classical system
of oscillators it was shown \cite{Caruso:2011tu} that this analogy can be
promoted to an equivalence between the Schroedinger dynamics of a quantum
system with a finite number of basis states and classical dynamics. The
equivalence is an isomorphism that connects in univocal way both dynamical
systems. There have been several other proposals and realizations
\cite{Rosner2},\cite{Kostelecky-Roberts},\cite{Reiser:2012fh},\cite{Schubert:2011hd},\cite{Caruso:2016kpy}
of the equivalence between the physics of the neutral K-meson quantum system
and classical systems of oscillators either electrical or mechanical. In
particular, in Ref.\ \cite{Rosner2}, Rosner and Slezak presented the emulation
of CP violation in the kaon system provided by the two-dimensional motion of a
Foucault pendulum. In their conclusions one reads: ``the classical emulation of
CP violation via the Foucault pendulum leaves us with one big puzzle.'' This
statement is because for this emulation they need to consider an asymmetry in
the coupling between the two modes of the pendulum, that classically is imposed
by the rotation of Earth. In the case of kaon mixing, that effect is ``thought
to arise from a complex phase in the CKM matrix.''  The authors finally asked
themselves if that phase is an indication of a new fundamental asymmetry from
beyond the standard model (SM).

Motivated by this comment we analyze in this note the connection between the
complex phase present in CP-violating systems such as the quantum one of
neutral kaons, and a geometrical phase such as the latitude effect acquired by
the Foucault pendulum. Moreover, we prove that the complex phase of the CKM
matrix, origin of the CP-violation in the framework of the SM, is a Berry-like
phase \cite{Berry}.

As is well known, the Berry phase appears when a quantum
system evolves and returns to its initial physical state in an adiabatic way.
After this evolution, it acquires a memory of this evolution in the form of a
time-independent geometrical phase in the quantum wave function.

As we have stated, our analysis originates in the Foucault pendulum as an
emulation of the CP-violating system \cite{Rosner2}. The motion of the
Foucault pendulum acquires a latitude effect in the form of a phase shift
\cite{latitude}. This phase was shown \cite{Fphase} to be no more than an
example of the classical version of the Berry phase, namely the Hannay phase
\cite{Hannay}.

Let us remember the dynamics of the Foucault pendulum, both from the standard
way based upon a non-inertial reference frame and also in the way that the
Hannay phase is explicitly shown to appear.

The equation of motion of the Foucault pendulum related to the kaon system
needs slightly different natural frequencies in both directions of motion and
vastly different damping in these directions in order to emulate the
characteristics of neutral kaons \cite{Rosner2}.  Consequently the mass matrix
corresponding to the Foucault pendulum satisfying CPT invariance
($\mathcal{N}_{21} = -\mathcal{N}_{12}$) reads \cite{Rosner2}
\begin{equation}
\mathcal{N}=
\left( \begin{array}{cc}
 \omega_1 - \imath \gamma_1 & \imath\Omega' \\
-\imath\Omega' & \omega_2 - \imath \gamma_2
\end{array} \right)~.
\end{equation}
Here $\Omega'$ is in general complex.

The matrix for the kaon system, written in terms of the effective Hamiltonian
(Eq.\ (53) of Ref.\cite{Caruso:2011tu}) is
\begin{equation}
\mathcal{H}=
\left( \begin{array}{cc}
 \Sigma + \Delta\,\cos\alpha  & \imath\Delta\,\sin \alpha \\
-\imath\Delta\,\sin \alpha & \Sigma - \Delta\,\cos\alpha
\end{array} \right)
\end{equation}
where
\begin{equation}
\Sigma  = \frac{\mu_S + \mu_L}{2} = \frac{m_S - m_L - \imath(\gamma_S -
\gamma_L)}{2}\,\,\,\,\,; \,\,\,\,\,\Delta  = \frac{\mu_S - \mu_L}{2}
= \frac{m_S - m_L + \imath(\gamma_S - \gamma_L)}{2}
\end{equation}
with $m_S$ and $m_L$ being the short and long $K^0$-masses respectively and
$\gamma_S$, $\gamma_L$ the corresponding decay rates.  Consequently,
 \begin{equation}\label{op}
 \Omega' = \Delta\, \sin \alpha~;
 \end{equation}
 the phase $\alpha$ is given by
\begin{equation}
\exp(\imath \alpha) = \frac{1-\epsilon}{1+\epsilon} \,\,\,\,\Rightarrow\,\,\,\,
\alpha \simeq 2 \,\imath \,\epsilon
\end{equation}
where $\epsilon$ is the traditional parameter that measures CP violation in
mixing.

From Eq.\ (\ref{op}) one finds, after simple algebra and using the fact that
$\epsilon$ is very small ($\sim 10^{-3}$), that
\begin{eqnarray}
\Re (\Omega') & = & \frac{m_L -m_S}{2}\,\sin (2\Im (\epsilon))~,
\label{kaonfoucoult1}\\
\Im (\Omega') & = & \frac{\gamma_L -\gamma_S}{2}\,\sin (2\Im (\epsilon))~.
\label{kaonfoucoult2}
\end{eqnarray}
This $\Omega'$ plays the role of $\Omega = \Omega_E\,\sin \theta$ in the simple
Foucault pendulum. In this last expression, $\Omega_E$ is the Earth daily-rotation frequency. 
It includes the latitude effect on the pendulum motion measured by the latitude of the observation point, 
$\theta$. This effect is the
manifestation of the Hannay geometrical phase.  Now, the inclusion of
dissipation, always present for kaons, leads to a complex $\Omega'$. The real
part, being proportional to the mass difference, is related to the Hermitian
part of the Hamiltonian, while the imaginary part, proportional to the decay
rates, comes from its non-Hermitian part \cite{Garrison}. Due to the
approximation introduced by the smallness of $\epsilon$, the phase is the same
for both parts of $\Omega'$.

At this point it is worth remembering that the latitude effect we have already
mentioned refers to the fact that after the Earth has completed one day's
revolution, the plane of oscillation of the Foucault pendulum will not complete
a $2 \pi$ rotation. There is a ``defect'' in the angle of rotation of the pendulum plane measured by $\sin \theta$, with $\theta$ the
latitude where the pendulum is maintained. The analysis of the Foucault
pendulum's motion can be done without using the so-called fictitious forces
appearing in non-inertial frames. In fact \cite{Fphase} one can develop a
geometrical model of the pendulum in which it keeps a constant direction of
oscillation manifested as a parallel displacement along a curved surface of the
Earth with a rotation frequency much lower that the oscillation one. As we said
before, this is an example of the Hannay phase \cite{Hannay}. Then, one can
conclude that in our case of the kaon system, $(2\Im \epsilon)$ plays exactly
the role of the latitude, the geometrical phase.

This result can be rephrased by saying that the Earth's motion is to the
Foucault pendulum as the weak interaction Hamiltonian (the cause of the
splitting of masses between $K_S$ and $K_L$ and the differences between mean
lives) is to the kaon system. Moreover, the geometrical phase related to the
latitude phase is directly related to the $\epsilon$ parameter that measures
the CP-violation in kaons.

Let us now go to the quantum analysis of the Berry phase in our case of
interest --- neutral kaons. This analysis starts at the quark level where the
CKM matrix is active.

In the Berry analysis one starts with a quantum Hamiltonian that depends on a
real parameter, $R$, that is adiabatically varied (it changes with a frequency
that is much smaller than any other frequency present in the problem).

In our case of the kaon system, certainly this small frequency characterizing
the adiabatic change is the weak interaction part of the Hamiltonian, $H_W$,
that is a small perturbation to the strong interaction Hamiltonian whose scale
is the $K^0$ mass. On the other hand, the $K^0 - \bar{K^0} -K^0$-oscillations
provide the cycle that enters in the Berry analysis. In order to perform the
analysis we refer first to quarks and their charged current interactions
measured by the CKM matrix. In entering the explicit calculation to show our
assertions, let us remember the general structure of the CKM matrix:
\begin{equation}
CKM =
\left( \begin{array}{ccc}
 V_{ud} & V_{us} & V{ub} \\
V_{cd} & V_{cs} & V_{cb} \\
V_{td} & V_{ts} & V_{tb}
\end{array} \right)~,
\end{equation}
where each element of the matrix measures the vertex of both quarks involved
with the $W$ vector boson. As was proposed in Ref.\ \cite{brannen}
we consider a sequence of quantum weak transitions connected with a SM
contribution to the $K^0 - \bar{K^0}$ oscillation: namely, the sequence of
Fig.\ 1, that is the cycle $d\,\rightarrow\,c\,\rightarrow\,s\,
\rightarrow\,u\,\rightarrow\,d$.
\begin{figure}[htb]
\centerline{\includegraphics[scale= 1.0]{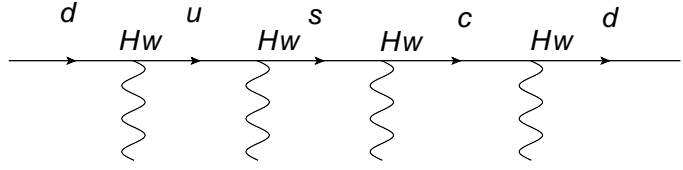}} \vskip 1cm \caption{One of
the sequences of weak vertices considered} \label{Fig1}
\end{figure}
The chosen sequence means that one has
\begin{eqnarray}
& &|d><d|H_W|u><u|H_W|s><s|H_W|c><c|H_W|d> \nonumber \\
&  & = \, V_{ud}\,V^\ast_{us}\,V_{cs}\,V^\ast_{cd}\,|d>~.
\end{eqnarray}
Note that we have maintained only the CKM contribution to each vertex.  In the
same way, now for the sequence $s\,\rightarrow\,u\,\rightarrow\,d\,
\rightarrow\,c\,\rightarrow\,s$, one obtains
\begin{eqnarray}
& & |s><s|H_W|u><u|H_W|d><d|H_W|c><c|H_W|s> \nonumber \\
&  &  = \,V_{us}\,V^\ast_{ud}\,V_{dc}\,V^\ast_{cs}\,|s>~.
\end{eqnarray}
Both previous equations are the ones related to the way quarks behave in the
$K^0 \rightarrow \bar{K^0} \rightarrow K^0$ oscillation and in the $\bar{K^0}
\rightarrow K^0 \rightarrow \bar{K^0}$ one, respectively. In order to go from
quarks to kaons, one has to add the contributions of the antiquark cycles, that
are
\begin{eqnarray}
& & |\bar{s}><\bar{s}|H_W|\bar{c}><\bar{c}|H_W|\bar{d}><\bar{d}|H_W|\bar{u}><\bar{u}|H_W|\bar{s}> \nonumber \\
&  &  = \, V_{cs}\,V^\ast_{cd}\,V_{ud}\,V^\ast_{us}\,|\bar{s}>~,
\end{eqnarray}
and
\begin{eqnarray}
& & |\bar{d}><\bar{d}|H_W|\bar{c}><\bar{c}|H_W|\bar{s}><\bar{s}|H_W|\bar{u}><\bar{u}|H_W|\bar{d}> \nonumber \\
&  & =  \, V_{cd}\,V^\ast_{cs}\,V_{us}\,V^\ast_{ud}\,|\bar{d}>~,
\end{eqnarray}
respectively.

Consequently, the amplitude for the cyclic oscillation $K^0-\bar{K^0}-K^0$ is
proportional to
\begin{equation}
\left( V_{ud}\,V^\ast_{us}\,V_{cs}\,V^\ast_{cd} \right)^2 =(A+\imath B)^2
\end{equation}
while for the cyclic oscillation $\bar{K^0}-K^0-\bar{K^0}$ the amplitude is proportional to
\begin{equation}
\left( V_{us}\,V^\ast_{ud}\,V_{cd}\,V^\ast_{cs}\right)^2 = (A-\imath B)^2~,
\end{equation}
which is exactly the complex conjugate of the previous one.  Clearly, the
difference between the two previous expressions is proportional to the amount
of CP-violating effects of the weak interaction.  It results in
\begin{equation}
2\,A\,B = \Im\left(V_{ud}\,V^\ast_{us}\,V_{cs}\,V^\ast_{cd} \right) = J_{CP}~,
\end{equation}
the Jarlskog invariant measuring CP violation \cite{Jar}.

Notice that $J_{CP}$, being invariant, does not depend (except for a sign) on
the pair of quantum states chosen. Moreover, $\epsilon$ is related to
$J_{CP}$ by \cite{thomson}

\begin{equation}
|\epsilon| \propto A_{ut}\, \Im{(V_{ud} V^\ast_{us} V_{td} V^\ast_{ts})} +
A_{ct}\, \Im{(V_{cd}V^\ast_{cs}V_{td} V^\ast_{ts})} + A_{tt}\,
\Im{(V_{td}V^\ast_{ts}V_{td} V^\ast_{ts})}~,
\end{equation}
where the imaginary parts are related to the Jarlskog invariant and the $A$'s
are numerical coefficients arising from  integrations over virtual momenta.

Recalling Eqs.\ (\ref{kaonfoucoult1}) and (\ref{kaonfoucoult2}) we unveil the
geometric character of the CKM matrix elements through the Jarlskog invariant.

In summary, we have shown that the complex phase of the CKM matrix, responsible
for the CP violation present in the SM, is of geometrical origin, entirely
similar to the quantum Berry phase. This result opens some new roads in the
analysis and possible measurements of CP-violation effects.

\section*{Acknowledgments}
We thank Jonathan Rosner for a very careful reading of the manuscript which has led to precise comments and suggestions that
have  improved the presentation. 
This work has been partially supported by ANPCyT and CONICET of Argentina, and
by MINECO (Spain) Grants No.\ FPA2013-47443-C2-1-P, GVA-PROMETEOII/2014/066,
and SEV-2014-0398.


\begin{thebibliography}{99}

\bibitem{Rosner} J. L. Rosner, \textit{Table top time reversal violation}, Am.\
J. Phys.\ \textbf{64} (8), 982-985 (1996).

\bibitem{Caruso:2011tu}
M. Caruso, H.~Fanchiotti and C.~A.~Garcia Canal, \textit{Equivalence between
classical and quantum dynamics. Neutral kaons and electric circuits},
 Ann.\ Phys.\ {\bf 326}, 2717 (2011) (arXiv:1104.1968 [quant-ph]).

\bibitem{Rosner2}
J. L. Rosner and S. A. Slezak, \textit{Classical
illustrations of CP violation in kaon decays}, Am.\ J. Phys.\ \textbf{69},
44-49 (2001).

\bibitem{Kostelecky-Roberts}
V. A. Kostelecky, A. Roberts, \textit{Analogue
models for T and CPT violation in neutral-meson oscillations}, Phys.\ Rev.\ D
\textbf{63}, 096002 (2001).

\bibitem{Reiser:2012fh}
A.~Reiser, K.~R.~Schubert and J.~Stiewe,
\textit{Translation of time-reversal violation in the neutral K-meson system
into a table-top mechanical system}, J.\ Phys.\ G {\bf 39}, 083002 (2012)
(arXiv:1203.4703 [hep-ph]).

\bibitem{Schubert:2011hd}
K.~R.~Schubert and J.~Stiewe, \textit{TOPICAL REVIEW:
Demonstration of ${\rm K^0}{\overline{\rm K}}{}^0$, ${\rm B^0}
{\overline{\rm B}}{}^0$ and ${\rm D^0}{\overline{\rm D}}{}^0$ transitions
with a pair of coupled pendula}, J.\ Phys.\ G {\bf 39}, 033101 (2012)
(arXiv:1108.2772 [hep-ph]).

\bibitem{Caruso:2016kpy}
M.~Caruso, H.~Fanchiotti, C.~G.~Canal, M.~Mayosky and V.~Veiga,
 \textit{ The quantum CP-violating kaon system reproduced in the electronic
laboratory (Homage to Nolberto Martinez)}
  Proc.\ Roy.\ Soc.\ Lond.\ A {\bf 472}, no. 2195, 20160615 (2016).

 \bibitem{Berry} M. V. Berry,
  \textit{Quantal Phase Factors Accompanying Adiabatic Changes},
  Proc.\ Roy.\ Soc.\ Lond.\ A {\bf 392} (1802), 45 (1984).

 \bibitem{latitude} W. B. Somerville,
 \textit{The Description of Foucault's Pendulum},
 Q.\ Jl.\ Astr.\ Soc.\ {\bf 13}, 40 (1972).

  \bibitem{Fphase} J.~B.~Hart, R.~E.~Miller and R.~L.~Miller,
 \textit{A simple geometric model for visualizing the motion of a Foucault
 pendulum}, Am.\ J.\ Phys.\ {\bf 55}, 67 (1987);
A. Khein and D. F. Nelson,
\textit{Hannay angle study of the Foucault pendulum in action-angle variables},
  Am.\ J.\ Phys.\ {\bf 61}, 170 (1993).

 \bibitem{Hannay} J. H. Hannay,
 \textit{Angle variable holonomy in adiabatic excursion of an integrable
Hamiltonian}, J. Phys.\ A {\bf 18}, 221 (1985).

\bibitem{Garrison} J. C. Garrison and E. M. Wright,
 \textit{Complex Geometrical Phases for Dissipative Systems},
 Phys.\ Lett.\ A {\bf 128}, 177 (1988).

 \bibitem{brannen} C.~A.~Brannen,
 \textit{Unitary mixing matrices and their parametrizations},
(arXiv: 1511.0083) (2012).

 \bibitem{Jar} C. Jarlskog, \textit{Commutator of the Quark Mass Matrices in
the Standard Electroweak Model and a Measure of Maximal CP Nonconservation},
Phys.\ Rev.\ Lett.\ {\bf 55}, 1039 (1985).

\bibitem{thomson} M. Thomson, \textit{Modern Particle Physics},
Cambridge University Press (2013).

\end{thebibliography}
\end{document}